# Benchmark tests of atom segmentation deep learning models with a consistent dataset

Jingrui Wei[a], Ben Blaiszik[b,c], Aristana Scourtas[b,c], Dane Morgan[a], and Paul M. Voyles[a]*

[a]Department of Materials Science and Engineering, University of Wisconsin-Madison, Madison, Wisconsin, United States

[b]Globus, University of Chicago, Chicago, Illinois, United States

[c]Data Science and Learning Division, Argonne National Laboratory, Lemont, Illinois, United States

*paul.voyles@wisc.edu

**Abstract**

The information content of atomic resolution scanning transmission electron microscopy (STEM) images can often be reduced to a handful of parameters describing each atomic column, chief amongst which is the column position. Neural networks (NNs) are a high performance, computationally efficient method to automatically locate atomic columns in images, which has led to a profusion of NN models and associated training datasets. We have developed a benchmark dataset of simulated and experimental STEM images and used it to evaluate the performance of two sets of recent NN models for atom location in STEM images. Both models exhibit high performance for images of varying quality from several different crystal lattices. However, there are important differences in performance as a function of image quality, and both models perform poorly for images outside the training data, such as interfaces with large difference in background intensity. Both the benchmark dataset and the models are available using the Foundry service for dissemination, discovery, and reuse of machine learning models.

## 1. Introduction

The information content of atomic-resolution electron microscopy images is often contained in the precise position, integrated intensity, width, and shape of each of the atomic columns within the image. That information can be used for studies of defects (Lee et al. 2020), dynamics in solid-state reactions (Pennycook et al. 2014), phase transformations (Maksov et al. 2019), inference of physical fields (Jones et al. 2017; Lim et al. 2019), and many other purposes. 2D Gaussian fitting (Bals et al. 2006) at the local maxima positions in an image has been the primary method used to estimate atomic column positions with picometer precision. Postprocessing methods like non-rigid registration (Yankovich et al. 2014) and denoising algorithms (Salmon et al. 2012; Zhang and Xia 2018; Zhang et al. 2020) also help achieve high precision by improving the signal-to-noise ratio of the original images and reducing distortion. However, these methods typically require manual hyperparameter tuning, substantial data redundancy in the form of an image series, and significant computing time, making them inefficient for problems requiring large numbers of columns or high speed.

Deep neural networks (DNNs) (LeCun et al. 2015) consist of sequential non-linear modules and can approximate unknown functions to accurately recognize patterns in data after training using data already labeled with the appropriate patterns. Applications of DNNs in image recognition have advanced from whole-image classification (Krizhevsky and Hinton 2012) to structured output like object detection (Sermanet et al. 2013 Dec 21), and then to pixel-level segmentation. Fully convolutional neural networks (FCNs) (Shelhamer et al. 2017) have been reported to outperform other convolutional networks at accurate segmentation of complex images. An FCN takes an arbitrary-sized image as input and produces a coarse feature map through several combinations of convolution, pooling, and activation layers, in a processing called downsampling or 'convolution'. A symmetrical upsampling or 'deconvolution' part, in which the pooling layers are replaced by bilinear interpolation layers, upsamples the coarse feature output back to the input image size. An FCN was first implemented on microscopy images by Ronneberger *et al.* for cell tracking (Ronneberger et al. 2015). Roberts *et al.* developed a network called DefectSegNet to classify crystallographic defects in the diffraction contrast STEM images of structural alloy (Roberts et al. 2019), and Lee *et al.* trained a network to locate and identify defects in averaged high-resolution STEM images and to enable strain mapping (Lee et al. 2020). Other types of NN have also been used to segment STEM images and classify crystallographic defects, including Faster Regional Convolutions Neural Network (Faster RCNNs) (Shen, Li, Wu, Liu, et al. 2021), Mask RCNN (Jacobs et al. 2022), and You Only Look Once (YOLO) (Shen, Li, Wu, Yaguchi, et al. 2021).

Here, we examine the performance of two FCNs developed by Lin *et al.* (Lin et al. 2021) and Ziatdinov *et al.* (Ziatdinov et al. 2017) for identifying atomic column positions in raw atomically resolved STEM images which may include significant noise, distortions, and poor contrast. The two networks have distinct network structure, several pre-trained model weights for users to choose from for different purposes, and they depend on different training datasets. Lin provides 15 different trained models for different tasks like atom segmentation and detection, noise reduction, background removal, and super-resolution processing. Ten of the models employ the basic U-Net architecture (Ronneberger et al. 2015), and the other five models use a nested U-Net structure as shown in **Figure 1(a)**. The models can also be used in combination. For example, Lin suggests first using one for noise/background removal, and then a second model for atom segmentation, to improve precision. The activation function used after convolution layers is rectified linear unit (ReLU) and the loss metric of training is a modified $\chi^2$ function, which reduces the false-positive rate better than the mean square error loss function. The FCN models give the probability map of being an atom column or background after a final classification layer (tanh or sigmoid). The probability map is binarized with Otzu's method (Otsu 1979) and the atom positions are calculated as the geometric centers of the disconnected area.

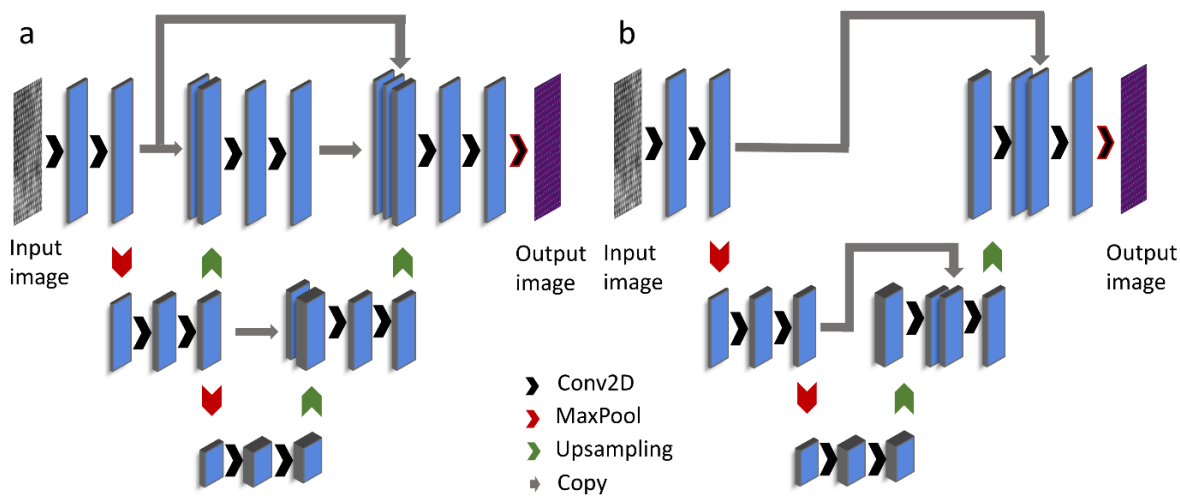

**Figure 1**. Unique FCN architecture design of the models. (a) The schematic architecture of the nested U-Net used by five of Lin's provided models: *Gen1-circularMask, Gen1-gaussianMask, Gen1-noBackgroundNonoise, Gen1-noNoise, Gen1-noNoiseNoBackgroundSuperresolution.* (b) The schematic architecture of FCN with residual connections used by Ziatdinov.

Ziatdinov provides two trained models, one for cubic lattices and the other for graphene-like lattices (Ziatdinov et al. 2017). Both models adopt residual blocks (He et al. 2016) that help preserve gradient norms during backpropagation and avoid bad local minima (Zaeemzadeh et al. 2021)(Kawaguchi and Bengio 2019) as shown in **Figure 1(b)**. The *cubic* model has a flatter net structure than Lin's, which means fewer steps in the downsampling / upsampling path but more stacked layers in each convolutional block. A special feature of the *graphene* model is dilated convolution at the bottleneck instead of regular convolution. These models make binary classification with a sigmoid function at the final layer and enable multi-classification using a softmax function (Duan et al. 2003). The atomic column coordinates are extracted via Laplacian of Gaussian blob detection on the probability map. The activation function is LeakReLU and the Adam optimizer is used with categorical cross-entropy as the loss function.

More recently, Ziatdinov *et al.* released a Pytorch-based python software for machine learning analysis of microscopy data called AtomAI (Maxim et al. 2020; S. V. Kalinin et al. 2021; S. V Kalinin et al. 2021). For atom segmentation tasks, AtomAI provides a FCN trainer so users can train a custom neural network specifically for their experimental data without much pre-knowledge of machine learning. This is a different approach than the general-purpose models described above, so we have investigated its performance as well.

Although these models and many other machine learning models for microscopy and other materials applications have been made freely available for reuse, few of them have been widely adopted outside their original research groups. We suggest at least two reasons for the limited adoption so far, although this is necessarily highly speculative. First, each model tends to be developed to solve a specific research problem, leaving its applicability on data outside the original test dataset unknown and difficult to evaluate. Further, the time and effort required to prepare a labeled dataset for a new task and evaluate the performance of an existing model can be significant, and it could be wasted if the model turns out to be unsuitable. Second, simply reusing the models from their current published form can be non-trivial. Models are usually published as code and weight archives on a site like Github, which in principle means they can be reused. In practice, reusing an existing model even a year or two old can require substantial effort if underlying libraries have been updated or other supporting software is no longer available.

These challenges to reuse of machine learning models are not unique to electron microscopy or even to materials science. The Foundry project (Blaiszik et al. 2022), leveraging the Materials Data Facility (Blaiszik et al. 2016; Blaiszik et al. 2019) and DLHub services (Chard et al. 2019), aims to promote reuse of models by mitigating both challenges. Foundry combines containerized ML models with their structured training data and metadata into a single digital object accessible through a python API. Foundry models preserve their runtime environment, including dependencies, in containers (e.g., Docker or Singularity) which can be executed on Foundry-provided computing, commercial or government cloud computing, or local computing resources with a single command. Foundry models also preserve their training data, including metadata, labels, and splits for training and validation, all of which is again accessible through simple commands. Foundry provides search functions for model and data discovery and DOIs for permanent reference, and both models and data can be updated to create new versions while preserving old versions in perpetuity. Thus, with just a few lines of code, and the associated DOIs, researchers can obtain the required datasets and access and run models without lengthy and error-prone installation processes. These features make Foundry models exceptionally Findable, Accessible, Interoperable, and Reusable (FAIR) (Wilkinson et al. 2016), helping researchers meet evolving standards in open science.

We have developed a benchmark dataset of experimental and simulated atomic-resolution STEM images and a set of model performance metrics for testing atom-locating approaches. The dataset is organized into image series based on seven image quality metrics, covering experimental images, simulated images, and a subgroup of various types of crystal lattices. We have evaluated the performance of the Lin and Ziatdinov models against the benchmark dataset using a set of performance metrics. Both models show excellent performance over different atomic lattices, although there are exceptions for each model. The key factor that limits a model's performance is the quality of STEM images, as a function of quality metrics including the pixel size, contrast, and image resolution. The data and models are all available on Foundry and the supplemental information includes example Jupyter notebooks using the Foundry interface. We hope these results and the Foundry interface will encourages others to reproduce our results, benchmark their own FCNs or other methods against the same data, augment the training dataset with new examples (especially additional experiments!), and train new, better, and more general models for STEM data.

## 2. Methods

**Benchmark dataset.** Table 1 summarizes the contents of the benchmark dataset. The dataset consists of 300 simulated and experimental images, all labeled with the atomic column positions. Simulations provide distortion-free, noiseless images for which the ground-truth atomic column positions are known. The simulated images can be degraded separately and quantitatively over a wide variety of image qualities. Experimental images are the ultimate targets for the models. Ground truth atomic column positions of experimental images are not known *a priori*, so we have manually labeled the atomic column positions in each image, then refined them using 2D Gaussian fitting to determine an approximate ground truth. The images were simulated or collected in part to varying a set of image quality metrics, as summarized in Table 1.

**Table 1**: Benchmark dataset contents. The dataset consists of 300 images which systematically vary a set of image quality metrics for two different lattices, $SrTiO_3$ [100] and $WS_2$ [0001], and images of various other lattices, defects, and interfaces.

|  |  | Simulated |  | | Experimental | | |
|---|---|---|---|---|---|---|---|
| *Quality Metric* | *Method* | *$SrTiO_3$ [100]* | *$WS_2$ [0001]* | *Method* | *$SrTiO_3$ [100]* | *$WS_2$ [0001]* | *Other lattices* |
| Pixel size (pm) | Simulation probe step size, image resize | 7-60 | 5-50 | Microscope magnification | 10.88-43.82 | 5.44-15.46 | - |
| Resolution (pm) | Gaussian blur | 113.1-329.8 | 106.3-300.8 | Gaussian blur | 103.0-317.2 | 93.5-295.0 | - |
| Contrast (%) | Gain and black level | 3.0-32.4 | 4.-37.2 | Gain and black level | 3.4-29.6 | 3.7-15.9 | - |
| Electron Dose ($e^-$/ Å) | - | - | - | Dwell time | ~1200-30000 | ~2900-35000 | - |
| Flyback error (pixel) | Interpolation | 0.1-2.9 | 0.5-3.9 | - | - | - | - |
| Random Drift (pixel) | Interpolation | 0.04-0.26 | 0-0.30 | - | - | - | - |
| Poisson Noise (PSNR dB) | Add Poisson noise | 1.36-22.77 | 0.10-20.58 | - | - | - | - |
| Total number of images |  | 86 | 99 |  | 37 | 32 | 46 |

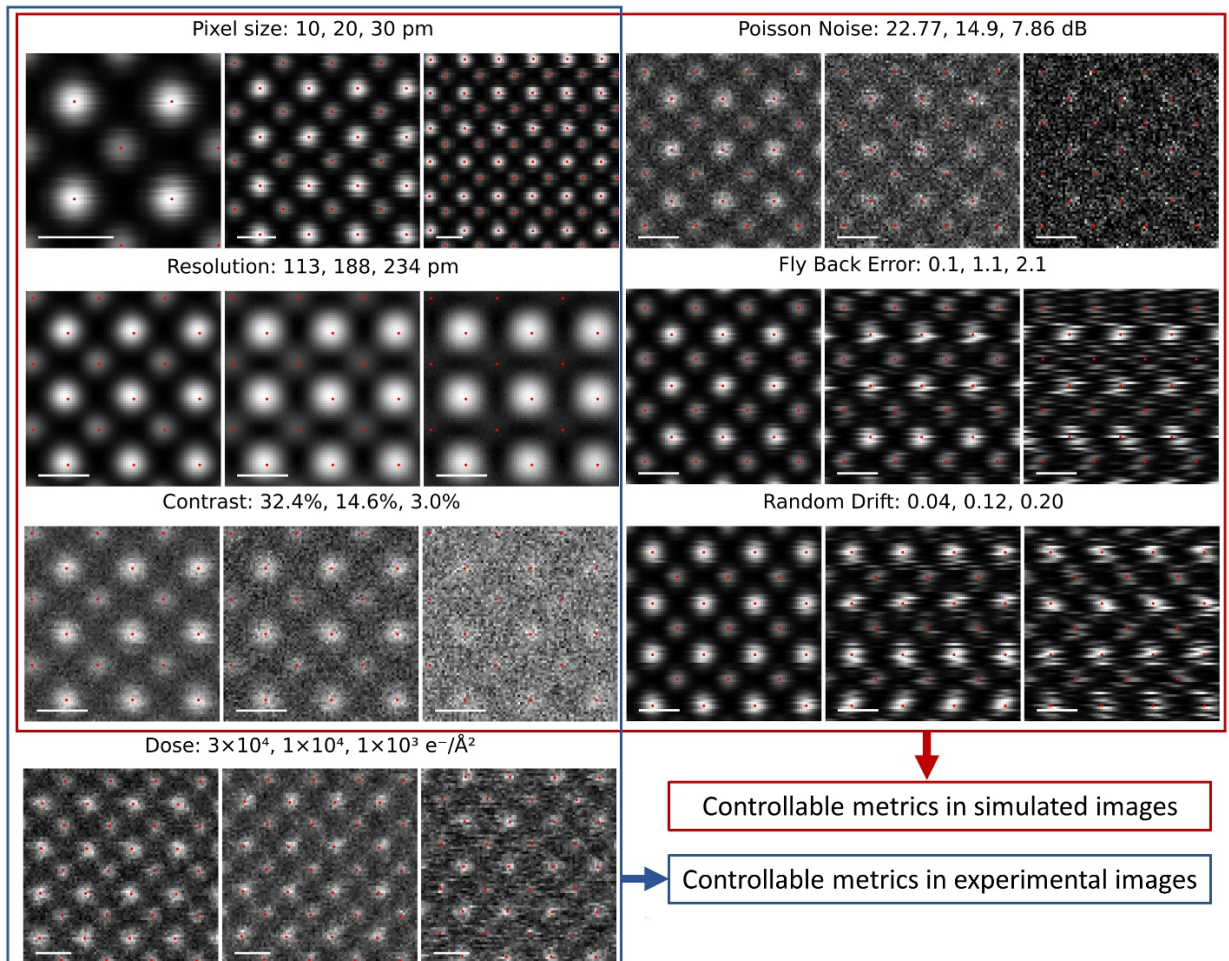

**Figure 2**. Example of image series for each quality metric on SrTiO$_3$. Some metrics can be varied only in simulations or experiments, and some can be varied in both. Ground truth positions of Lin's model are marked as red dots. Scale bars represent 3 Å.

Images of SrTiO$_3$ [100] and two-dimensional (2D) WS$_2$ [0001] are the major part of the dataset, and are both included in the training data for Lin's and Ziatdinov's models. The dataset also contains 46 experimental images covering a wide variety of crystal lattices and orientations, complex unit cells, heterointerfaces, grain boundaries, and thickness variation. These images are intended to test the general applicability of the models on "typical" high-resolution STEM data, so they are not restricted to the lattices in the training data for the models.

To evaluate the model performance, the dataset includes a series of images as a function of several different quality metrics. For simulated images, six series of images were generated separately as a function of pixel size, resolution, contrast, Poisson noise, scan flyback error, and random sample drift. For experimental images, four series of images were generated separately as a function of pixel size, resolution, contrast, and electron dose. **Figure 2** shows the poor quality, normal quality, and high-quality images for each metric. Each image series only varies in one quality metric, and the rest of the metric values remain constant at a typical level for experiments. Pixel size is adjusted in simulations by changing the probe step size and image resampling and in experiments by changing the microscope magnification. Resolution is adjusted in simulations by blurring with a Gaussian kernel with different standard deviations, mimicking different incoherent source sizes on the microscope. Contrast is adjusted in simulations by changing the gain factor and black level, which most simulates how a HAADF detector modifies contrast. Poisson noise is adjusted in simulations using a Poisson distributed random number generator to achieve noise characterized by the peak signal-to-noise ratio (PSNR) (Horé and Ziou 2010). Probe flyback error and random sample drift are adjusted in simulations by random shifts of scan lines or the entire scan grid, respectively, with respect to perfect image simulation grid, then interpolation. The displacements are drawn randomly from normal distributions and the distortion is varied by changing the standard deviation (Berkels and Wirth 2017).

**Table 2.** Multislice simulation parameters. Thermal displacement parameters are taken from Ref. (Zuo and Spence 2016) (Appendix F).

|  | SrTiO$_3$ [100] | WS$_2$ [0001] |
| --- | --- | --- |
| Lattice volume | 1 × 1 × 64 unit cells | 40 × 40 × 1 unit cells |
| Probe step size | 0.10 Å | 0.20 Å – 0.50 Å |
| Potential sampling | 0.06 Å | 0.06 Å |
| Frozen phonons | 20 | 20 |
| Root mean square thermal displacement | Sr: 0.0887 Å, Ti: 0.0746 Å, O: 0.0947 Å | W: 0.0230Å, S: 0.0710 Å |

We have formatted this benchmark dataset into a queryable, interoperable, and reusable HDF5 file. One group is created for each of the 300 images, containing a floating-point raw image array, an 8-bit image for easy visualization, ground-truth x and y coordinates of the atom columns in units of pixels, and metadata that describes the image source, image scale, crystal lattice, value of quality metrics, probe current, *etc.* The dictionary-like structure of the HDF5 file allows users to select desired images by specifying filter criteria based on the metadata, making this dataset shapeable for multiple purposes.

We generated the simulated STEM images of 10 nm thick SrTiO$_3$ and bilayer WS$_2$ using the multislice algorithm in the GPU-accelerated Prismatic simulation package (Pryor et al. 2017). The thermal vibration of atoms was considered using the Debye-Waller factor at room temperature. Table 2 gives a full list of simulation parameters. Experimental HAADF STEM images are acquired on the FEI Titan microscope at University of Wisconsin-Madison equipped with a CEOS probe aberration corrector operated at 200 keV. They are collected from current and former research from the Voyles Group and generally use a probe semi-angle between 23.4-24.5 mrad and a probe current between 18.9-39.4 pA.

**Model evaluation.** We use three metrics to evaluate the performance of the models, ID-recall, ID-precision, and position deviation, $\Delta d$. ID-recall measures the fraction of ground truth atoms that have been correctly located and is defined as (TP/(FN+TP)), where TP is the number of true positive atom column identifications, and FN is the number of false negatives. ID-precision measures the fraction of model predictions that are correct and is defined as (TP/(FP+TP)), where FP is the number of false positives. We define a TP, *i.e.* a correctly predicted position, as a distance between the predicted position and the ground-truth position smaller than 0.5 Å. A FP is a predicted position that is farther than 0.5 Å from any ground truth column, and a FN is a ground truth atom column with no predicted position within 0.5 Å. We set the criteria at 0.5 Å as it less than 60% of the smallest atomic column spacing in the test dataset. A slightly different criteria would result in different evaluation scores, but the comparison results based on relative score value remain the same. $\Delta d$ is the root mean square distance between the TP model-identified positions and the corresponding ground truth positions.

**Figure 3** illustrates the procedure we employed for determining a quantitative cutoff as a function of image quality, using contrast as an example. For quality past this cutoff, we consider a model to be inaccurate for practical use. We evaluate the model performance using ID-recall, ID-precision, and $\Delta d$ as a function of image quality, then define cutoffs at the image quality past which at least one of the following occurs: ID-recall falls below 90%, ID-precision falls below 95%, or position deviation exceeds 0.3 Å. Based on the observation of multiple test cases and common atomic column spacings, we believe these three criteria need to be satisfied to extract accurate information from atomically resolved images. Again, a slightly different criteria would result in different evaluation scores, but the comparison results based on relative score value remain the same. The cutoffs are calculated from sigmoid function fit or pchip interpolation, and the model performance is reported as the best image quality at the cutoff for the three metrics. **Figure 3(a-c)**, show that as contrast decreases, Lin's model exhibits falling ID-recall because it misses atoms, while ID-Precision and position deviation remain at good levels. **Figure 3(d-f)** shows that for Ziatdinov's model, ID-Precision falls at higher contrast than ID-Recall because the model generates many false positives at low contrast. For Ziatdinov's models, the model maintains good performance for contrast of 0.114 or higher for experiments and 0.172 for experiments.

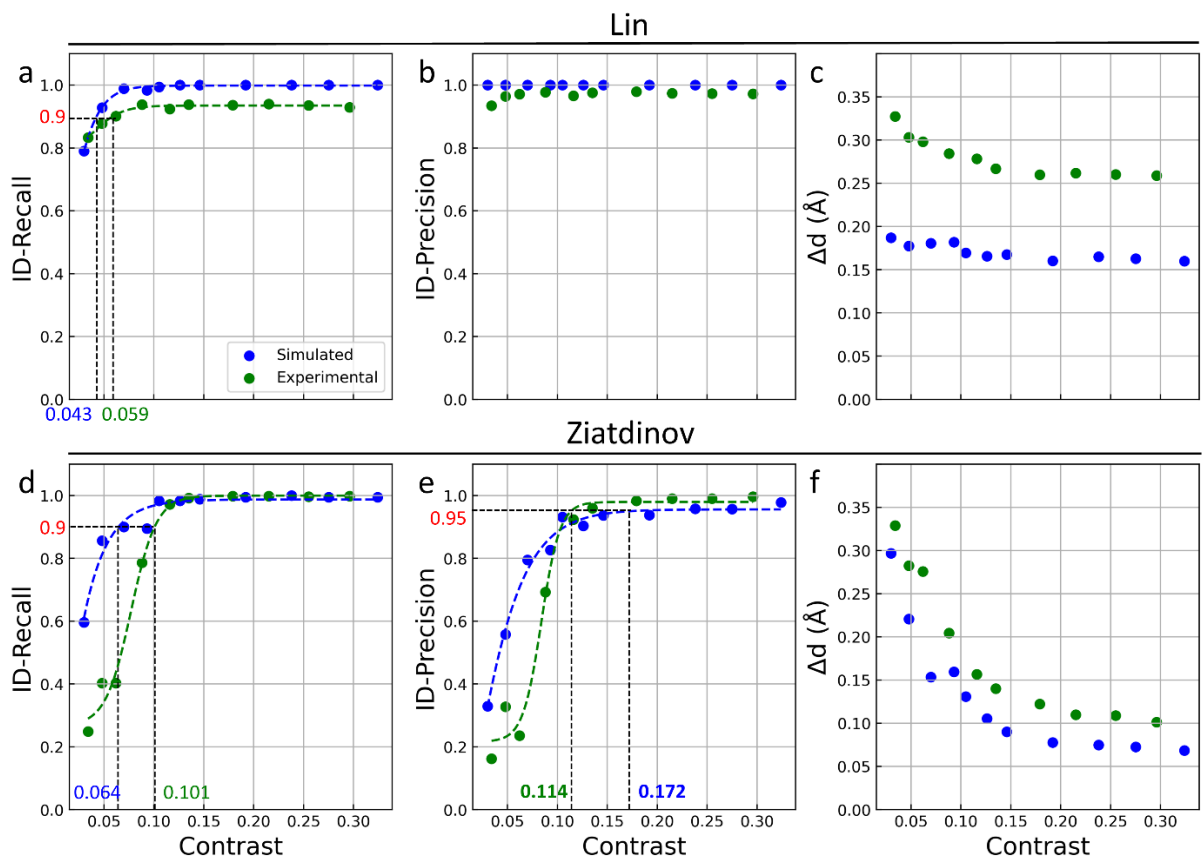

**Figure 3**. Examples of determining performance cutoffs as a function of the contrast of SrTiO$_3$ simulated and experimental images. (a-c) The three evaluation scores for Lin's model, resulting in a contrast cutoff of **NN**. (d-f) The three evaluation scores for Ziatdinov's model.

We tested all 15 of Lin's models and found that the *GaussianMask+* model and the *Gen1-noNoiseNoBackgroundSuperresolution* model had the best performance, so those results are presented throughout. We did not test all of the possible combinations, on the principle that an ideal tool would not require preprocessing (*e.g.* denoising) from the user. Therefore, better performance using more complex workflows tuned for specific images or conditions may be possible. For Ziatdinov's models, we applied the graphene model on hexagonal lattices like WS$_2$ and the *cubic* model on all other lattices, including SrTiO$_3$.

To evaluate the quality of new models created with the AtomAI package, we trained several new models using its default training hyperparameters, listed in **Table 3**. The models were trained on simulated images of specific crystal lattices for different comparisons described below. The test data are either simulated images or experimental images of typical quality and always separate from the training dataset.

**Table 3.** AtomAI training hyperparameters

| Data augmentation | (Scaled units) | Training | |
|---|---|---|---|
| Background | True | Network architecture | SegResNet (Ziatdinov et al.) |
| Rotation | True | Number of classes | 1 |
| Zoom | 1 | Loss function | Cross entropy |
| Jitter | 0-5 | Optimizer | Adam |
| Poisson | 20-55 | Learning rate | 0.001 |
| gauss | 20-90 | Batch size | 32 |
| blur | 1-100 | Batch normalization | True |
| contrast | 5-18 | Drop out | False |
| salt_and_pepper | 1-20 | Dataset size | 300 ~ 1500 image patches with 256×256 pixels |

### 3. Results and Discussion

### 3.1 Performance against image quality

**Figure 4** summarizes the performance of the two models using radar plots of the cutoffs for all explored image quality metrics. In these plots, points further away from the center mean that the model maintains acceptable performance at poorer image quality, so in general a larger area polygon is "better". The units of pixel size were scaled to atomic columns per pixel as a consistent metric for lattices with different atomic density. **Figure 4(a, c)** shows the cutoffs for Lin's models and Ziatdinov's *cubic* model on SrTiO$_3$ [100] images. Ziatdinov's model shows acceptable performance on lower quality images except for contrast. Lin's model is more restricted by the pixel size of the image. **Figure 4(b, d)** shows the cuttoffs for Lin's models and Ziatdinov's *graphene* model on 2D WS$_2$ [0001] images. Ziatdinov's model shows relatively better performance, especially for pixel size and resolution. The cutoffs for experimental images are always lower than those from simulated images, but comparison between models is consistent. The limitation on pixel size for both models can be mitigated by resizing the images via interpolation if the other quality metrics are high enough. The larger polygon of Ziatdinov's models in **Figures 4(b)** and **(c)** also may imply that models trained on a single type of lattice outperform the models like Lin's that are trained on many different lattices. In other words, there may be a trade-off between model performance against image quality and general applicability. However, at the outermost edges of the radar plots, the image quality is so poor that it is difficult to discern the atomic columns by eye. Thus, these tests demonstrate the overall power of the FCN models for atomic column detection, even for very poor images.

**Figure 5** compares the performance of three models trained on SrTiO$_3$ images, then tested on other SrTiO$_3$ simulations as a function of image quality. One model is a SegResNet model trained using AtomAI (Ziatdinov et al.). The second is the Ziatdinov *cubic* model, retrained using AtomAI. The third is the Ziatdinov *cubic* model itself, reproduced from Figure 4(a) for comparison. The new AtomAI model (green polygon) has good performance when applied to images with the same or similar lattice type of its training data, but it does not outperform the *cubic* model. Retraining the *cubic* model to just the SrTiO$_3$ lattice improves performance over several image quality metrics, but reduces the pixel size range to the range covered in the training data.

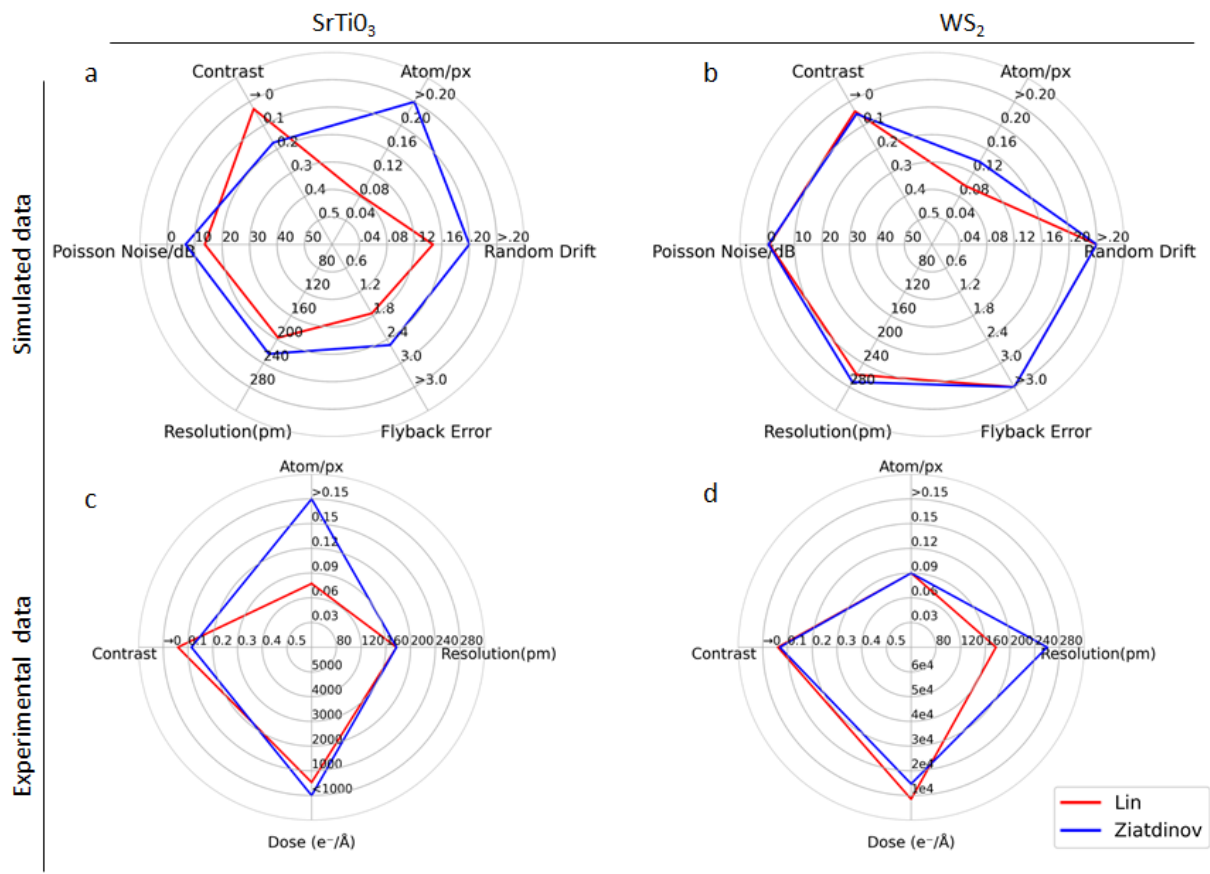

**Figure 4**. Visualization of strengths and limitations of the two models by radar plots as a function of image quality metrics. (a) Tests on simulated $SrTiO_3$ data, (b) tests on simulated $WS_2$ data, (c) tests on experimental $SrTiO_3$ data, (d) tests on experimental $WS_2$ data.

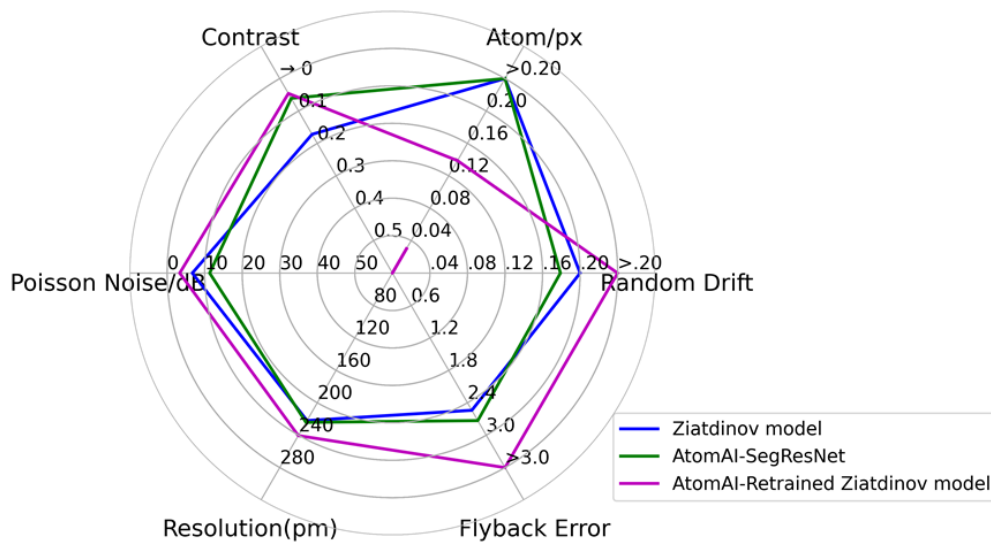

**Figure 5**. Performance of a new model trained with AtomAI, Ziatdinov's *cubic* model retrained with AtomAI, and Ziatdinov's original *cubic* model, all tested on $SrTiO_3$ simulated images.

## 3.2 General applicability

There is as yet no general approach to quantify the domain of a trained deep learning model, so we have taken a purely empirical approach to determining which images fall in the domains of applicability of the tested models. Here we investigate the applicability of Lin and Ziatdinov model using 46 different experimental images. Figure 6(a)-(c) show the ID-recall, ID-precision, and Δd, which were optimized by resizing the images. Generally down-sampling improves ID-Precision and up-sampling improves ID-recall. For the most part, the ID-recall and ID-precision are very good. The Δd is 5-25 pm, which is within the range reported for standard fitting methods on single-shot STEM images (Schmid et al. 2012). Figure 6(d)-(i) show example images for which the performance is poor. Low SNR, especially for dim atomic columns, is the common cause of poor performance of Lin's model as shown in **Figure 6(d, e)**. Ziatdinov's model has poor performance in various ways, including detecting many false positives in the background (**Figure 6f**), missing many atoms if the image contains significant background intensity variation (**Figure 6g**), failing to distinguish overlapping atomic columns (**Figure 6h**), and showing poor performance for images with poor resolution or severe distortion (**Figure 6i**).

We emphasize that these tests are a deliberate attempt to cause the models to fail. They are in that sense unfair, and they are not how the model's creators would recommend using the models. Despite that, the model performance remains quite high for a wide variety of images. However, whether a particular image will cause the models to perform poorly remains difficult to predict, so the models should be used for structures outside the training datasets with caution.

AtomAI suggests broadening applicability of FCN methods by training a new model for each set of similar images. Simulated training images can be used to reduce the effort of obtaining and hand-labeling training data. AtomAI simplifies the workflow of training dataset preparation, network assembling, and model validation, in order to build a quick path to a customized model. We used AtomAI to train an FCN model on simulated Si [110] images and tested it on the GaAs [110] lattice image. The performance is significantly improved compared to the Ziatdinov *cubic* model, with an ID-recall of 100% compared with 62.1%.

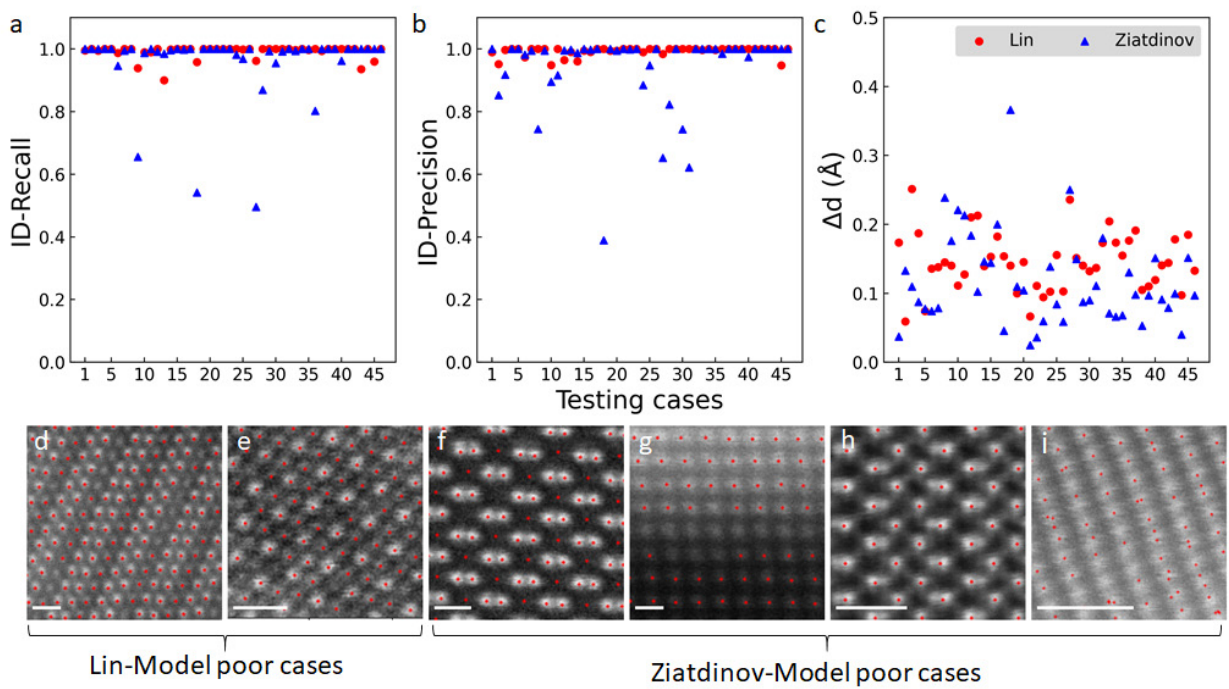

**Figure 6.** The general applicability of Lin's model and Ziatdinov's model represented by tests on 46 different experimental images. (a-c) Best evaluation scores of the two models after resampling, (d,e) typical poor test cases for Lin's model, (f-h) typical poor test cases for Ziatdinov's model. The red dots mark model identified atom positions. Scale bars represent 3 Å.

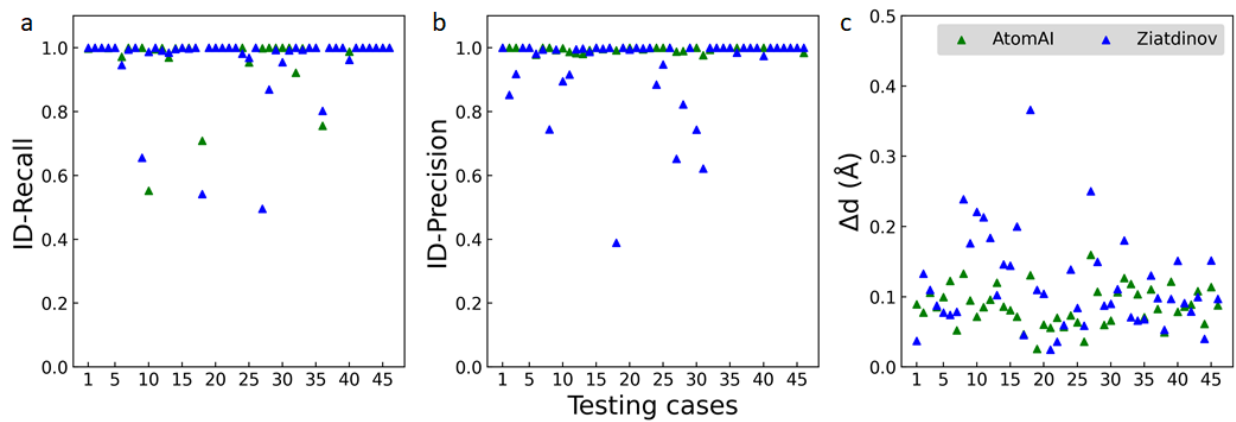

**Figure 7**. General applicability comparison of the model trained on five crystal lattice types using AtomAI and Ziatdinov model. (a) ID-recall, (b) ID-Precision, (c) Δd for each of the 46 test cases.

However, one general model for any atomic-resolution image remains a desirable goal. We carried out a cross-validation-like test to investigate the possibility of applying AtomAI to create such a generalized model with a modest amount of training data. Simulated images of five crystal lattice types, $SrTiO_3$ [100], $SrTiO_3$ [110], Si [110], $WS_2$ [0001], and $DyScO_3$ [110], were augmented as described in **Table 3** to create a dataset. Models were trained on images of four lattices, then validated on the fifth lattice. The ID-recall and ID-precision of all five models on the validation data were all 100%. Further tests on experimental $SrTiO_3$ [110], $WS_2$ [0001], and GaN [110] images of fair quality retain good performance, with an averaged ID-recall of 97.8%, ID-precision of 100%, and Δd of 16 pm.

**Figure 7** compares the performance on our 46 experimental generality test images of an AtomAI model trained on the simulations of all five crystal lattices with Ziatdinov model. The AtomAI-trained model generates few false positives and deals better than Ziatdinov model with images with smoothly varying background intensity. Its typical cases with low ID-recall comes from images of poor resolution, overlapping atomic columns, and interfaces with significant jumps in intensity. These results suggest that an FCN trained on only five crystal lattice types is already more general than the current specialized models.

Exploring broader training datasets, possibly incorporating generative adversarial networks to mimic experimental noise and distortion, to create a more general network is an interesting direction for future research (Lee et al. 2020). The Foundry service facilitates this kind of research by making datasets and models readily available, executable, and FAIR. All of the datasets discussed here are available through Foundry (Wei et al. 2021; Wei, Blaiszik, et al. 2022), which includes ground truth labels, structured metadata, and python methods for accessing subsets of the data, all for easy reuse of data for testing other models or training new models. Foundry also offers version control and incorporates citations so a new dataset adding to on one of ours can be submitted while maintaining appropriate credit. All the models discussed here can be executed in Foundry containers (Lin et al. 2022; Wei, Ziatdinov, et al. 2022a; Wei, Ziatdinov, et al. 2022b; Wei, Ziatdinov, et al. 2022c; Wei, Ziatdinov, et al. 2022d; Wei, Ziatdinov, et al. 2022e; Wei, Ziatdinov, et al. 2022f; Wei, Ziatdinov, et al. 2022g; Wei, Ziatdinov, et al. 2022h; Ziatdinov 2022), and an example Jupyter notebook of retrieving the datasets and executing models can be found in the supporting information.

## 4. Conclusions

We have tested the performance of two sets published FCNs designed to identify atomic column positions in high-resolution STEM micrographs using a benchmark dataset of simulated and experimental STEM images. The benchmark dataset contains several series of images with systematic variation in different measurements of image quality like pixel size and signal to noise ratio, and a set of experimental images of different crystal lattices, defects, and interfaces to evaluate the general applicability of models. We also explored using the neural network training framework AtomAI to create new models designed for specific problems and for general use.

In general, all of the models do very well at identifying atom column positions in a wide variety of high-resolution STEM images, including images in which the columns are essentially invisible to the human eye due to poor quality and images outside the training datasets for the models. The accuracy of the positions is typically 10-20 pm. Different models varying in their performance envelope for images with large pixels, low contrast, or other forms of poor quality. All the models have poor performance on at least some of the generality test images, but the specific image feature that result in poor performance remain difficult to predict *a priori*. In general, Lin's model has strong performance for images with poor contrast and Ziatdinov's models have good performance on images with other forms of poor quality. For images in which both models perform poorly, perhaps because they contain complex features like dim or overlapping atomic columns or an interface, a custom model trained using AtomAI may be a viable alternative. The benchmark dataset and the models are published using the Foundry service to provide easy, fast, and FAIR access and hopefully accelerate progress in atom segmentation approaches for STEM.

## Acknowledgments

This work was primarily supported by the National Science Foundation project "Collaborative Research: Framework: Machine Learning Materials Innovation Infrastructure" (UW-Madison: OAC-1931298, U. Chicago OAC-1931306), including development of the benchmark dataset, evaluation of the neural network model performance, and implementation of the Foundry model and data dissemination system. Additional support for infrastructure and development associated with the Materials Data Facility is provided by project 70NANB14H012 of the U.S. Department of Commerce, National Institute of Standards and Technology as part of the Center for Hierarchical Material Design (CHiMaD). Experiments were performed using facilities supported by the Wisconsin MRSEC (DMR-1728933).

## References

Bals S, Van Aert S, Van Tendeloo G, Ávila-Brande D. 2006. Statistical estimation of atomic positions from exit wave reconstruction with a precision in the picometer range. Phys Rev Lett. 96(9):1–4. doi:10.1103/PhysRevLett.96.096106.

Berkels B, Wirth B. 2017. Joint denoising and distortion correction of atomic scale scanning transmission electron microscopy images. Inverse Probl. 33(9):1–35. doi:10.1088/1361-6420/aa7b94.

Blaiszik B, Chard K, Pruyne J, Ananthakrishnan R, Tuecke S, Foster I. 2016. The Materials Data Facility: Data Services to Advance Materials Science Research. JOM. 68(8):2045–2052. doi:10.1007/s11837-016-2001-3. https://doi.org/10.1007/s11837-016-2001-3.

Blaiszik B, Scourtas A, Schmidt K. 2022. Foundry. https://github.com/MLMI2-CSSI/foundry.

Blaiszik B, Ward L, Schwarting M, Gaff J, Chard R, Pike D, Chard K, Foster I. 2019. A data ecosystem to support machine learning in materials science. MRS Commun. 9(4):1125–1133. doi:10.1557/mrc.2019.118. https://doi.org/10.1557/mrc.2019.118.

Chard R, Li Z, Chard K, Ward L, Babuji Y, Woodard A, Tuecke S, Blaiszik B, Franklin MJ, Foster I. 2019. DLHub: Model and Data Serving for Science. In: 2019 IEEE International Parallel and Distributed Processing Symposium (IPDPS). p. 283–292.

Duan K, Keerthi SS, Chu W, Shevade SK, Poo AN. 2003. Multi-category Classification by Soft-Max Combination of Binary Classifiers. In: Windeatt T, Roli F, editors. Multiple Classifier Systems. Berlin, Heidelberg: Springer Berlin Heidelberg. p. 125–134.

He K, Zhang X, Ren S, Sun J. 2016. Deep residual learning for image recognition. Proc IEEE Comput Soc Conf Comput Vis Pattern Recognit. 2016-Decem:770–778. doi:10.1109/CVPR.2016.90.

Horé A, Ziou D. 2010. Image quality metrics: PSNR vs. SSIM. Proc - Int Conf Pattern Recognit.(August):2366–2369. doi:10.1109/ICPR.2010.579.

Jacobs R, Shen M, Liu Y, Hao W, Li X, He R, Greaves JRC, Wang D, Xie Z, Huang Z, et al. 2022. Performance and limitations of deep learning semantic segmentation of multiple defects in transmission electron micrographs. Cell Reports Phys Sci. 3(5):100876. doi:10.1016/J.XCRP.2022.100876.

Jones L, Wenner S, Nord M, Ninive PH, Løvvik OM, Holmestad R, Nellist PD. 2017. Optimising multi-frame ADF-STEM for high-precision atomic-resolution strain mapping. Ultramicroscopy. 179:57–62. doi:10.1016/j.ultramic.2017.04.007.

Kalinin S V., Dyck O, Jesse S, Ziatdinov M. 2021. Exploring order parameters and dynamic processes in disordered systems via variational autoencoders. Sci Adv. 7(17):1–10. doi:10.1126/sciadv.abd5084.

Kalinin S V, Zhang S, Valleti M, Pyles H, Baker D, De Yoreo JJ, Ziatdinov M. 2021. Disentangling Rotational Dynamics and Ordering Transitions in a System of Self-Organizing Protein Nanorods via Rotationally Invariant Latent Representations. ACS Nano. 15(4):6471–6480. doi:10.1021/acsnano.0c08914. https://doi.org/10.1021/acsnano.0c08914.

Kawaguchi K, Bengio Y. 2019. Depth with Nonlinearity Creates No Bad Local Minima in ResNets. Neural Netw. 118:167–174.

Krizhevsky A, Hinton GE. 2012. ImageNet Classification with Deep Convolutional Neural Networks. F. Pereira, C. J. C. Burges, L. Bottou KQW, editor. Curran Associates, Inc.

LeCun Y, Bengio Y, Hinton G. 2015. Deep learning. Nature. 521(7553):436–444. doi:10.1038/nature14539. https://doi.org/10.1038/nature14539.

Lee CH, Khan A, Luo D, Santos TP, Shi C, Janicek BE, Kang S, Zhu W, Sobh NA, Schleife A, et al. 2020. Deep learning enabled strain mapping of single-atom defects in two-dimensional transition metal dichalcogenides with sub-picometer precision. Nano Lett. 20(5):3369–3377. doi:10.1021/acs.nanolett.0c00269.

Lim A, Zhang C, Strohbeen PJ, Shourov EH, Rodolakis F, Fredrickson DC, Kawasaki JK. 2019. High electrical conductivity in the epitaxial polar metals LaAuGe and LaPtSb High electrical conductivity in the epitaxial polar metals LaAuGe and LaPtSb. 121107(October). doi:10.1063/1.5132339.

https://doi.org/10.1063/1.5132339.

Lin R, Zhang R, Wang C, Yang X, Xin H. 2022. AtomSegNet. doi:10.26311/8S9H-DZ64. https://doi.org/10.26311/8S9H-DZ64.

Lin R, Zhang R, Wang C, Yang XQ, Xin HL. 2021. TEMImageNet training library and AtomSegNet deep - learning models for high - precision atom segmentation , localization , of atomic - resolution images. Sci Rep.:1–15. doi:10.1038/s41598-021-84499-w. https://doi.org/10.1038/s41598-021-84499-w.

Maksov A, Dyck O, Wang K, Xiao K, Geohegan DB, Sumpter BG, Vasudevan RK, Jesse S, Kalinin S V, Ziatdinov M. 2019. Deep learning analysis of defect and phase evolution during electron beam-induced transformations in WS2. npj Comput Mater. 5(1):12. doi:10.1038/s41524-019-0152-9. https://doi.org/10.1038/s41524-019-0152-9.

Maxim Z, Jesse S, Sumpter BG, Kalinin S V, Dyck O. 2020. Tracking atomic structure evolution during directed electron beam induced Si-atom motion in graphene via deep machine learning. Nanotechnology. 32(3):35703. doi:10.1088/1361-6528/abb8a6. http://dx.doi.org/10.1088/1361-6528/abb8a6.

Otsu N. 1979. A Threshold Selection Method from Gray-Level Histograms. IEEE Trans Syst Man Cybern. 9(1):62–66. doi:10.1109/TSMC.1979.4310076.

Pennycook TJ, Jones L, Pettersson H, Coelho J, Canavan M, Mendoza-Sanchez B, Nicolosi V, Nellist PD. 2014. Atomic scale dynamics of a solid state chemical reaction directly determined by annular dark-field electron microscopy. Sci Rep. 4(1):7555. doi:10.1038/srep07555. https://doi.org/10.1038/srep07555.

Pryor A, Ophus C, Miao J. 2017. A streaming multi-GPU implementation of image simulation algorithms for scanning transmission electron microscopy. Adv Struct Chem Imaging. 3(1):15. doi:10.1186/s40679-017-0048-z. https://doi.org/10.1186/s40679-017-0048-z.

Roberts G, Haile SY, Sainju R, Edwards DJ, Hutchinson B, Zhu Y. 2019. Deep Learning for Semantic Segmentation of Defects in Advanced STEM Images of Steels. Sci Rep. 9(1):12744. doi:10.1038/s41598-019-49105-0. https://doi.org/10.1038/s41598-019-49105-0.

Ronneberger O, Fischer P, Brox T. 2015. U-net: Convolutional networks for biomedical image segmentation. Lect Notes Comput Sci (including Subser Lect Notes Artif Intell Lect Notes Bioinformatics). 9351:234–241. doi:10.1007/978-3-319-24574-4_28.

Salmon J, Deledalle C-A, Willett R, Harmany Z. 2012. Poisson noise reduction with non-local PCA. In: ICASSP.

Schmid H, Okunishi E, Oikawa T, Mader W. 2012. Structural and elemental analysis of iron and indium doped zinc oxide by spectroscopic imaging in Cs-corrected STEM. Micron. 43(1):49–56. doi:10.1016/j.micron.2011.05.011.

Sermanet P, Eigen D, Zhang X, Mathieu M, Fergus R, Lecun Y. 2013 Dec 21. OverFeat: Integrated Recognition, Localization and Detection using Convolutional Networks. Int Conf Learn Represent.

Shelhamer E, Long J, Darrell T. 2017. Fully Convolutional Networks for Semantic Segmentation. IEEE Trans Pattern Anal Mach Intell. 39(4):640–651. doi:10.1109/TPAMI.2016.2572683.

Shen M, Li G, Wu D, Liu Y, Greaves JRC, Hao W, Krakauer NJ, Krudy L, Perez J, Sreenivasan V, et al.

2021. Multi defect detection and analysis of electron microscopy images with deep learning. Comput Mater Sci. 199(August). doi:10.1016/j.commatsci.2021.110576.

Shen M, Li G, Wu D, Yaguchi Y, Haley JC, Field KG, Morgan D. 2021. A deep learning based automatic defect analysis framework for In-situ TEM ion irradiations. Comput Mater Sci. 197:110560. doi:10.1016/J.COMMATSCI.2021.110560.

Wei J, Blaiszik B, Morgan D, Voyles P. 2021. Benchmark Dataset for Locating Atoms in STEM images. doi:10.18126/E73H-3W6N. https://petreldata.net/mdf/detail/foundry_wei_atom_locating_benchmark_v1.1.

Wei J, Blaiszik B, Morgan D, Voyles PM. 2022. Training Dataset for Locating Atoms in STEM images . doi:10.18126/QSDL-AJ6X. https://petreldata.net/mdf/detail/training_locating_atoms_stem_images_v1.2.

Wei J, Ziatdinov M, Voyles MP. 2022a. Atomai SegResNet Trained on 5 Crystal Lattices. doi:10.26311/E2MW-QF63. https://doi.org/10.26311/E2MW-QF63.

Wei J, Ziatdinov M, Voyles MP. 2022b. AtomAI SegResNet trained on Si[110]. doi:10.26311/B6ZB-NS88. https://doi.org/10.26311/B6ZB-NS88.

Wei J, Ziatdinov M, Voyles MP. 2022c. AtomAI SegResNet trained on STO[100]. doi:10.26311/Q6E2-2P11. https://doi.org/10.26311/Q6E2-2P11.

Wei J, Ziatdinov M, Voyles MP. 2022d. Atomai SegResNet_1 for 5-fold CV. doi:10.26311/BKK2-GC19. https://doi.org/10.26311/BKK2-GC19.

Wei J, Ziatdinov M, Voyles MP. 2022e. Atomai SegResNet_2 for 5-fold CV. doi:10.26311/K2BK-HW50. https://doi.org/10.26311/K2BK-HW50.

Wei J, Ziatdinov M, Voyles MP. 2022f. Atomai SegResNet_3 for 5-fold CV. doi:10.26311/BGB7-K519. https://doi.org/10.26311/BGB7-K519.

Wei J, Ziatdinov M, Voyles MP. 2022g. Atomai SegResNet_4 for 5-fold CV. doi:10.26311/X13G-7F17. https://doi.org/10.26311/X13G-7F17.

Wei J, Ziatdinov M, Voyles MP. 2022h. Atomai SegResNet_5 for 5-fold CV. doi:10.26311/S8HF-3V6. https://doi.org/10.26311/S8HF-3V6.

Wilkinson MD, Dumontier M, Aalbersberg IjJ, Appleton G, Axton M, Baak A, Blomberg N, Boiten J-W, da Silva Santos LB, Bourne PE, et al. 2016. The FAIR Guiding Principles for scientific data management and stewardship. Sci Data. 3(1):160018. doi:10.1038/sdata.2016.18. https://doi.org/10.1038/sdata.2016.18.

Yankovich AB, Berkels B, Dahmen W, Binev P, Sanchez SI, Bradley SA, Li A, Szlufarska I, Voyles PM. 2014. Picometre-precision analysis of scanning transmission electron microscopy images of platinum nanocatalysts. Nat Commun. 5(May):1–7. doi:10.1038/ncomms5155.

Zaeemzadeh A, Rahnavard N, Shah M. 2021. Norm-Preservation: Why Residual Networks Can Become Extremely Deep? IEEE Trans Pattern Anal Mach Intell. 43(11):3980–3990. doi:10.1109/TPAMI.2020.2990339.

Zhang A, Xia D. 2018. Tensor SVD: Statistical and Computational Limits. IEEE Trans Inf Theory. 64(11):7311–7338. doi:10.1109/TIT.2018.2841377.

Zhang C, Han R, Zhang AR, Voyles PM. 2020. Denoising Atomic Resolution 4D Scanning Transmission Electron Microscopy Data with Tensor Singular Value Decomposition. Ultramicroscopy. 219(September):113123. doi:10.1016/j.ultramic.2020.113123. https://doi.org/10.1016/j.ultramic.2020.113123.

Ziatdinov M. 2022. Ziatdinov AtomNet. doi:10.26311/BF7A-7071. https://doi.org/10.26311/BF7A-7071.

Ziatdinov M, Dyck O, Maksov A, Li X, Sang X, Xiao K, Unocic RR, Vasudevan R, Jesse S, Kalinin S V. 2017. Deep Learning of Atomically Resolved Scanning Transmission Electron Microscopy Images: Chemical Identification and Tracking Local Transformations. ACS Nano. 11(12):12742–12752. doi:10.1021/acsnano.7b07504.

Ziatdinov M, Ghosh A, Wong T, Kalinin S V. AtomAI: A Deep Learning Framework for Analysis of Image and Spectroscopy Data in (Scanning) Transmission Electron Microscopy and Beyond. http://energy.gov/downloads/doe-public-access-plan.

Zuo JM, Spence JCH. 2016. Advanced transmission electron microscopy: Imaging and diffraction in nanoscience.